\documentclass[aps,a4paper,showpacs,prc,floatfix,twocolumn,amsmath,amssymb,nofootinbib]{revtex4-1}
\usepackage{graphicx}
\usepackage{epsfig}
\usepackage{ulem}
\usepackage{morefloats}
\usepackage{rotating}
\usepackage{color,ulem}

\def\beq{\begin{equation}}
\def\eeq{\end{equation}}
\def\beqa{\begin{eqnarray}}
\def\eeqa{\end{eqnarray}}
\def\ban{\begin{eqnarray*}}
\def\ean{\end{eqnarray*}}
\def\bi{\begin{itemize}}
\def\ei{\end{itemize}}

\begin{document}

\title{Low-density in-medium effects on light clusters from heavy-ion data}
\author{Helena Pais$^1$}
\author{R\'emi Bougault$^2$}
\author{Francesca Gulminelli$^2$}
\author{Constan\c ca Provid{\^e}ncia$^1$}
\author{Eric Bonnet$^3$}
\author{Bernard Borderie$^4$} 
\author{Abdelouahad  Chbihi$^5$}
\author{John D. Frankland$^5$} 
\author{Emmanuelle Galichet$^{4,6}$}
\author{Di\'ego Gruyer$^2$}
\author{Maxime Henri$^5$}
\author{Nicolas Le Neindre$^2$} 
\author{Olivier Lopez$^2$}
\author{Loredana Manduci$^{2,7}$}
\author{Marian Parl\^og$^2$}
\author{Giuseppe Verde$^8$}

\affiliation{$^1$CFisUC, Department of Physics, University of Coimbra,
  3004-516 Coimbra, Portugal. \\
$^2$Normandie Univ., ENSICAEN, UNICAEN, CNRS/IN2P3, LPC Caen, F-14000 Caen, France. \\
$^3$SUBATECH UMR 6457, IMT Atlantique, Universit\'e de Nantes, CNRS-IN2P3, 44300 Nantes, France. \\
$^4$Universit\'e Paris-Saclay, CNRS/IN2P3, IJCLab, 91405 Orsay, France. \\
$^5$Grand Acc\'el\'erateur National d'Ions Lourds (GANIL), CEA/DRF-CNRS/IN2P3, Bvd. Henri Becquerel, 14076 Caen, France. \\
$^6$Conservatoire National des Arts et Metiers, F-75141 Paris Cedex 03, France. \\
$^7$Ecole des Applications Militaires de l'Energie Atomique, BP 19 50115, Cherbourg Arm\'ees, France. \\
$^8$INFN - Sezione Catania, via Santa Sofia 64, 95123 Catania, Italy.}

\begin{abstract}

The modification of the ground state properties of light atomic nuclei in the nuclear and stellar medium is addressed, using chemical equilibrium constants evaluated from  a new analysis of the intermediate energy heavy-ion (Xe$+$Sn) collision data measured by the INDRA collaboration. 
Three different reactions are considered, mainly differing by the isotopic content of the emission source.
The thermodynamic conditions of the data samples are extracted from the measured multiplicities allowing for a parametrization of the in-medium modification, determined with the single hypothesis that the different nuclear species in a given sample correspond to a unique common value for the density of the expanding source. 
We show that this correction, which was not considered in previous analyses of chemical constants from heavy ion collisions, is necessary, since the observables of the analyzed systems show strong deviations from the expected results for an ideal gas of free clusters.
This data set is further compared to a relativistic mean-field model, and seen to be reasonably compatible with a universal correction of the attractive $\sigma$-meson coupling.

\end{abstract}

\maketitle

Light nuclear clusters play an important role in the warm and low density nuclear matter \cite{Oertel2017,Hempel2017} that can be found in core-collapse supernovae (CCSN) and neutron star (NS) mergers \cite{Arcones2008,Sumiyoshi2008,Fischer2014,Furusawa2013,Furusawa2017}.
Their main role is to affect the weak interaction rates, and, as a consequence, the dynamic evolution of these violent events \cite{Oertel2017}. In NS mergers, their abundance  has a direct influence on the fraction of the ejecta that is converted into r-process elements \cite{Bauswein2013}, or on the viscous evolution of the accretion disk after the merger \cite{Rosswog2015}, and, therefore, on the amount of matter that becomes unbound from the disk.
 
The most popular equations of state (EoS) only consider  $\alpha$ clusters (for a list of available EoS see the CompOSE database \cite{compose}), but deuterons and tritons have been proven to be more abundant than free protons \cite{Hempel2012,Fischer2014,Sumiyoshi2008}, and a reliable estimation of the cluster abundances in the different thermodynamic conditions is needed.
 
Nuclear clusters have been measured in heavy-ion collisions, under similar thermodynamic conditions. In Refs.~\cite{QinPRL108,Hagel}, the authors presented  chemical equilibrium constants for four light clusters, that were the first available constraints for in-medium modifications of light clusters at finite temperature. 

The problem of evaluating those abundances arises from the fact that their ground state properties are expected to be modified in a dense medium. Mass shifts arising from in-medium correlations were calculated in the framework of quantum-statistical approaches, but only in a limited density domain, and for a limited number of nuclear species \cite{Roepke2015}. For this reason, phenomenological models were developed \cite{Typel2010}, where the interactions with the medium are governed by coupling constants that must be fixed through comparison with experiment \cite{HempelPRC91,PaisPRC97}.

Very recently \cite{BougaultJPG19}, the INDRA collaboration presented  new sets of data, complementing the unique constraint which was previously available from NIMROD data \cite{QinPRL108}. However, as already observed in Ref.~\cite{BougaultJPG19}, the weak point of both experimental works is that the thermodynamic parameters, in particular the  baryon density $\rho_B$ and the temperature $T$, are not directly measured, but they are deduced from the experimental multiplicities using analytical expressions that explicitly assume that the physical system under study can be modeled as an ideal gas of clusters. This is in contradiction with the very purpose of the analysis, which is to extract the in-medium modifications with respect to the ideal gas limit. 

In this Letter, we propose to solve this methodological inconsistency by modifying the ideal gas expressions relating the thermodynamical parameters to cluster yields. This correction is estimated using Bayesian techniques, under the unique condition that the volume associated to the thermal motion of each cluster species should be the same,  which is a necessary (even if not sufficient) condition to be able to interpret  the experimental sample in terms of thermodynamic equilibrium.  

The new chemical constants evaluated from the INDRA data \cite{BougaultJPG19} using this improved data analysis  are then compared to the relativistic mean field model of Ref.~\cite{PaisPRC97}, in order to extract the in-medium modifications.  
We show that a single parameter, expressing a universal reduction of the scalar attractive field to the nucleons bound in clustered states, can be tuned so as to obtain a reasonably good description of the  chemical constants. The suppression effect is smaller than the one  obtained from the comparison to the equilibrium constants of Ref.~\cite{QinPRL108}, where ideal gas expressions were used to extract the thermodynamical parameters, but still corresponds to important in-medium modifications of the binding energies.

Under well-defined thermodynamic conditions, as given by the temperature $T$, total baryon density $\rho_B$ and proton fraction $y_p$,  equilibrium chemical constants  $K_c (A,Z)$ of a cluster of mass (charge) number $A$ ($Z$), are defined in terms of he number of clusters per volume, i.e. the particle densities $\rho_{AZ}$, or of mass fractions $\omega_{AZ}$ as : 
\begin{eqnarray}
K_c (A,Z)=\frac{\rho_{AZ}}{\rho_{11}^{Z} \rho_{10}^{A-Z}}=\frac{\omega_{AZ}}{A \omega_{11}^{Z} \omega_{10}^{A-Z}}\rho_B^{-(A-1)} \, .  \label{eq:chemical}
\end{eqnarray}
An experimental measurement of such constants requires the detection of particles and clusters from a statistical ensemble of sources, and an estimation of the associated thermodynamic parameters $(T,\rho_B,y_p)$. 

Under the assumption that chemical equilibrium holds at the different time steps of the emission from the expanding source produced in central  $^{136,124}$Xe+$^{124,112}$Sn collisions, the Coulomb corrected particle velocity $v_{surf}$ in the source frame can be used to select  statistical ensembles of particles corresponding to different emission times, and therefore different thermodynamic conditions \cite{QinPRL108}. A detailed comparison between the four reactions was performed in Ref.~\cite{BougaultPRC97}, verifying the statistical character of the emission.  A strong argument confirming  the crucial hypothesis of chemical equilibrium as a function of time was given in Ref.~\cite{BougaultJPG19}, observing that the extracted thermodynamic parameters as a function of $v_{surf}$ are independent of the entrance channel of the reaction.

The detected multiplicities  $Y_{AZ}(v_{surf})$  allow a direct experimental determination of the mass fractions  $\omega_{AZ}=AY_{AZ}/A_T$, as well as of the total source mass $A_T(t)$ as a function of the emission time, but the measurement of the baryonic density $\rho_B(t)=A_T/V_T$ additionally requires an estimation of the source volume, at the different times of the expansion. 
This latter is given by the free volume $V_f$ with the addition of the proper volume $V_{AZ}$ of the clusters  which belong to the source at a given time, $V_T=V_f+\sum_{AZ}V_{AZ}{\omega_{AZ}A_T}/{A}$, with $V_{AZ} = 4\pi R_{AZ}^3 /3$, where $R_{AZ}$  is the experimental radius of each cluster.

The free volume can be extracted from the differential cluster spectra $\tilde Y_{AZ}(\vec {p})= Y_{AZ}(v_{surf})/(4\pi p^2 \Delta p) $, which can be related to differential cluster densities as $ f_{AZ}(\vec {p})={\tilde Y_{AZ}(\vec {p})}/{V_f}$ \cite{QinPRL108,BougaultJPG19}.  
Supposing an ideal gas of classical clusters with binding energies $B_{AZ}$  in thermodynamic equilibrium at temperature $T$ in the grand-canonical ensemble, the differential mass densities read:
\begin{equation}
 f^{id}_{AZ}(\vec {p})= \frac{g_{AZ}}{h^3}  \exp  \left [ \frac {1}{T} \left ( B_{AZ}- \frac{p^2}{2M_{AZ}} + 
Z\mu_p + N\mu_n \right )\right ], \label{eq:spectra}
\end{equation}
with $M_{AZ}=Am-B_{AZ}$, $g_{AZ}=2S_{AZ}+1$ the mass and spin degeneracy of cluster $(A,Z)$, $m$ the nucleon mass, and the superscript stands for ``ideal''.
In-medium effects are expected to suppress the cluster densities \cite{Roepke2015}, with respect to Eq.~(\ref{eq:spectra}), $\rho_{AZ}=C_{AZ}\rho_{AZ}^{id}$,
where the in-medium correction $C_{AZ}<1$ can depend on  the thermodynamic conditions, the cluster species and their momentum \cite{Roepke2015}.
If we normalize the cluster spectrum  by the proton and neutron spectra at the same velocity, the unknown chemical potentials $\mu_{n,p}$ cancel, and  the free volume $V_f$ can be independently estimated from the different cluster species as:

\begin{equation}
V_f
= h^3 R_{np}^{\frac{A-Z}{A-1}} C_{AZ} \exp \left [ \frac{B_{AZ}}{T(A-1)}\right ]\left (\frac {g_{AZ}}{2^A}\frac{\tilde Y_{11}^A(\vec p)}
{\tilde Y_{AZ}(A\vec p)} \right )^{\frac{1}{A-1}} \; , \label{eq:vf1}
\end{equation}
where the free neutron-proton ratio $R_{np}$ is estimated  from the multiplicities of the $A=3$ isobars ,
$R_{np}=\left({Y_{31}}/{Y_{32}}\right)\exp{\left[  ({B_{32} - B_{31}})/{T}\right]}$, and $B_{AZ}$ are the experimentally known vacuum binding energy of the clusters.

\begin{figure} [htbp]
  \begin{tabular}{cc}
\includegraphics[width=0.5\textwidth]{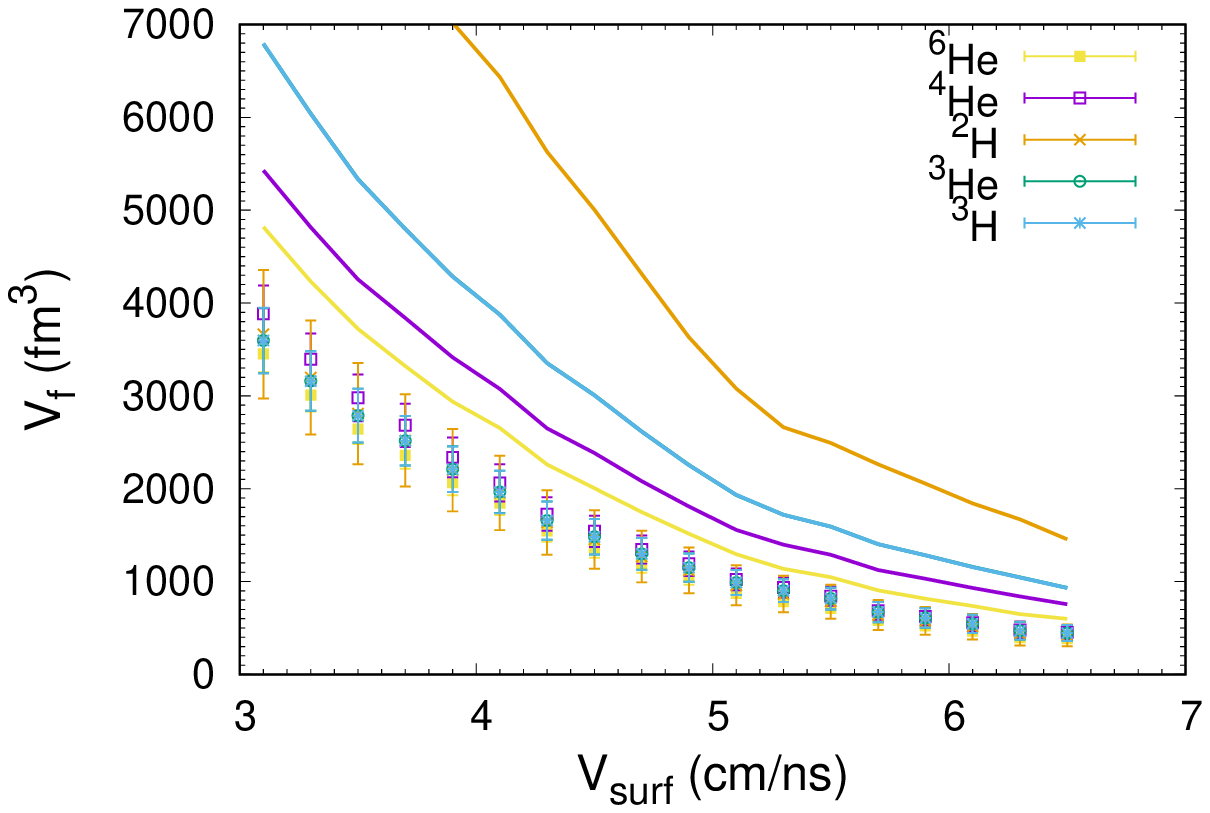}\\
\includegraphics[width=0.5\textwidth]{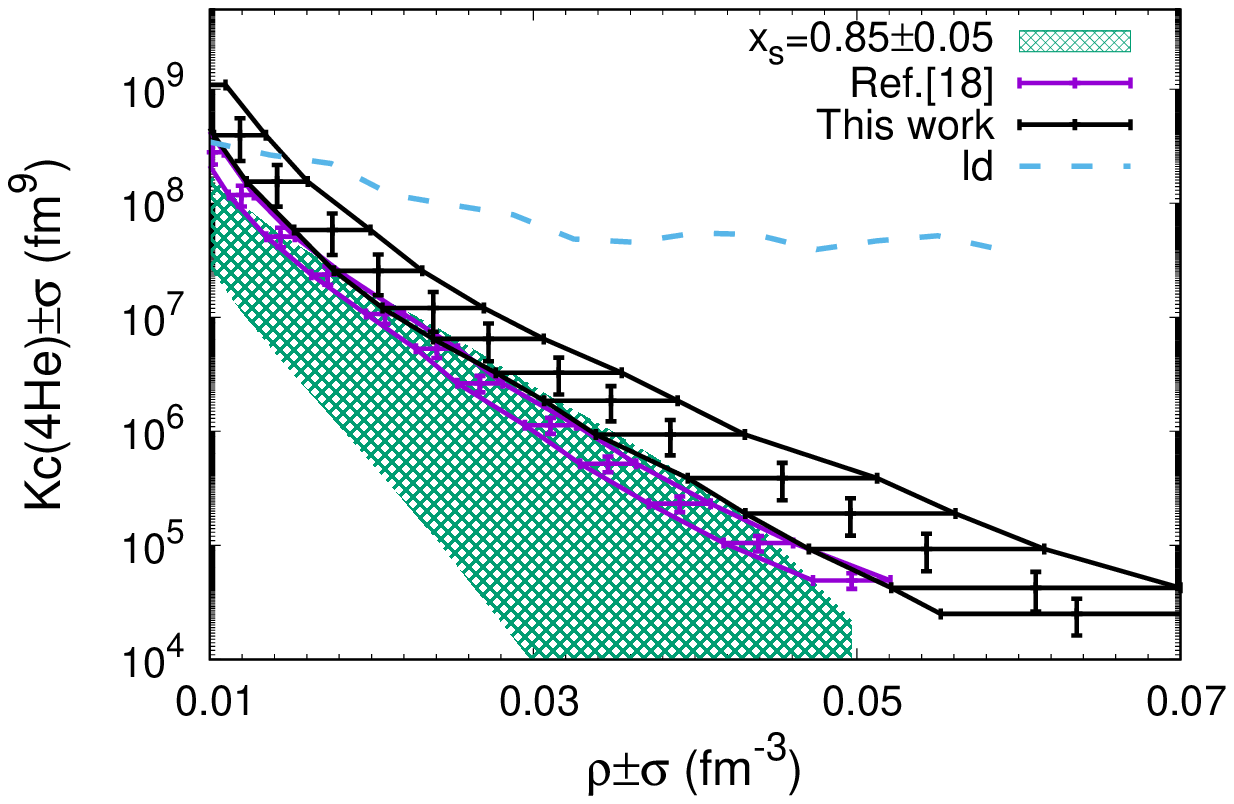}
  \end{tabular}
\caption{(Color online) System $^{124}$Xe$+^{112}$Sn. (Top) Free volume estimated from the different clusters as a function of $v_{surf}$ from Eq.~(\ref{eq:vf1}). Lines: ideal gas limit $C_{AZ}=1$. Note: the lines of $^3$H and $^3$He overlap. Symbols: bayesian determination of the in-medium correction. (Bottom) Chemical equilibrium constant of $^4$He as a function of the density, estimated from the data with the ideal gas prescription for the volume (lower set of points), and with the corrected one (upper set). For comparison, the predictions of Ref.~\cite{PaisPRC2019} with a coupling such as to fit the uncorrected results from Ref.~\cite{QinPRL108} are shown as a continuous band labelled $x_s=0.85\pm0.05$, and the ideal gas prediction is shown by a dashed line.}
\label{fig1}
\end{figure}  

The presence of in-medium corrections is clearly confirmed by the experimental data, as shown  by Fig.~\ref{fig1}, which displays the value of the free volume obtained from  Eq.~(\ref{eq:vf1}) for the $^{124}$Xe$+^{112}$Sn system, using different particle species.
A clear hierarchy is observed as a function of the cluster mass if $C_{AZ}=1$ is assumed, corresponding to the ideal gas limit. 
It is clear from Eq.~(\ref{eq:vf1}) that  to have consistent estimations of the volume, the deuteron requires a larger correction with respect to the heavier He isotopes. The volume splitting increases with decreasing $v_{surf}$, showing that the in-medium effects additionally depend on the thermodynamic conditions.
 Fully compatible results are obtained from the other three data sets (not shown).

The correction factors $C_{AZ}$ are, therefore, introduced as a modification of the cluster binding energies due to the presence of the medium. We introduce a very general four-parameters expression as:  

\begin{eqnarray}
C_{AZ}(\rho_B,y_p,T)&=& \exp \left [- \frac{a_1 A^{a_2}+a_3 |I|^{a_4}}{T_{HHe}(A-1)}\right ] \label{eq:correction} \, ,
\end{eqnarray}
where the temperature is estimated through the isobaric double isotope ratio Albergo formula \cite{AlbergoNCA89}, and it is indicated as $T_{HHe}$.
The unknown  parameters $\vec a =\{ a_i(\rho_B,y_p,T),i=1 - 4\}$ are taken as random variables, with a probability distribution fixed by imposing that the volumes obtained from the experimental spectra $\tilde Y_{AZ}$  of the different $(A,Z)$ nuclear species in a given $v_{surf}$ bin, correspond to compatible values.    To minimize the a-priori assumptions, we take in each $v_{surf}$ bin uninformative flat priors, $P_{\rm prior}(\vec a)=\theta(\vec a_{\rm min}-\vec a_{\rm max})$, within an interval largely covering the physically possible reduction range of the  binding energy,  $0\le a_1\le 15$ MeV, $0\le a_3\le a_1$, $-1\le a_2 \le 1$, $0 \le a_4 \le 4$.  

The posterior distribution is obtained by imposing the volume observation with a likelihood probability as follows:
\begin{eqnarray}
P_{post}(\vec a)={\cal N}\exp\left(-\frac{\sum_{AZ}(V_{f}^{(AZ)}(\vec a)-\bar V_f(\vec a))^2}{2\bar V_f(\vec a)^2}\right) \, . \label{eq:likely}
\end{eqnarray}  
Here, ${\cal N}$ is a normalization, $V_{f}^{(AZ)}(\vec a)$ is the free volume obtained from the $(A,Z)$ cluster using Eq.~(\ref{eq:vf1}) with the specific choice $\vec a$ for the parameter set of the correction, and $\bar V_f(\vec a)$ is the volume corresponding to a given parameter set $\vec a$, averaged over the cluster species.
 
The posterior expectation values of the volume as estimated from the multiplicities of each cluster from Eq.~(\ref{eq:vf1}), with the associated standard deviations, are shown as symbols in  Fig.~\ref{fig1}. 
It is clear that when we include the correction, the volumes decrease and the estimations obtained from the different cluster species  are compatible within error bars. 
Concerning the functional dependence of the correction, we can observe that we have as many parameters as different independent volume estimations, meaning that we are allowing independent corrections for the different nuclear species. It would be  interesting to have chemical constant measurements for other nuclear species, such as to check if a universal dependence of the in-medium effects on $A$ and $I$, as it is supposed in different theoretical models \cite{Typel2010,HempelPRC91,PaisPRC97}, is supported by the data. 

The bottom panel of Fig.~\ref{fig1} shows the corresponding modification of the $^4$He chemical equilibrium constant in the system  $^{124}$Xe$+^{112}$Sn. Similar results are obtained for the other particles and the other systems (not shown). In this Figure, the standard deviations associated to the experimental equilibrium constants are joined by full lines.  
The estimation with $C_{AZ}=1$ as in \cite{QinPRL108}, already shown in Ref.~\cite{BougaultJPG19}, is given by the lower set of points \footnote{It has to be noticed that the definition of chemical constants in Ref.~\cite{BougaultJPG19} differs by a factor $A$ with respect to the one of Refs.~\cite{QinPRL108,Roepke2015,PaisPRC97}. To allow an easier comparison with previous works, we have here adopted the definition of Ref.~\cite{QinPRL108}. Due to the different definitions, in Fig. 9 of Ref.~\cite{BougaultJPG19}, the NIMROD data should have been multiplied by a factor $A$ for a direct comparison.}, while the higher data set gives the result employing the posterior distribution of $C_{AZ}$ from Eq.~(\ref{eq:likely}).
We can see that both the average and the standard deviation of the estimation are increased.  
Concerning the effect on the average, a reduction of the volume corresponds to an increase of the baryonic density, up to a factor of two, and therefore an increase of the chemical equilibrium constants with respect to the estimation employing the ideal gas assumption (see Eq.~(\ref{eq:chemical})).
Concerning the variance, while in the previous analysis no experimental error was associated to the volume estimation, the bayesian determination of the volume distribution allows a more realistic estimation of the systematic uncertainties  of both density and chemical constants, with increased error bars. Realistic uncertainties might be even slightly larger on the low density side, because we cannot exclude that the in-medium effects could lead to an increased proper size of the clusters $V_{AZ}$.
The results of the different systems almost perfectly overlap, confirming the expectation that chemical constants are isospin-independent  (not shown).
 If we compare the experimental chemical constants with the ideal gas expectation Eq.~(\ref{eq:spectra}) (dashed line in Fig.~\ref{fig1}), we can observe an important suppression of $^4$He clusters at high density. But this suppression is less pronounced than the one obtained with the previous analysis, with important consequences on the present estimation of in-medium effects for theoretical applications in the astrophysical context, as we now discuss. 

In Ref.~\cite{PaisPRC97}, a novel approach for the inclusion of in-medium effects in the equation of state for warm stellar matter with light clusters was introduced. This model includes a phenomenological modification in the scalar cluster-meson coupling, and  includes an extra term in the effective mass of the clusters, which acts as an exclusion-volume effect.
The scalar coupling acting on nucleons bound in a cluster of mass $A$ is defined as  $g_{s}(A) = x_s A g_s$, with $g_s$ the scalar coupling of homogeneous matter, and $x_s$ a free parameter. A constraint on this parameter was obtained in the low-density regime from the Virial EoS, but a precise determination of $x_s$ needs an adjustment at densities close to the Mott density corresponding to the dissolution of clusters in the medium. The parameter $x_s$ measures how much the medium affects the binding of the cluster.
The smaller the $x_s$, the stronger the in-medium effect, and the smaller the dissolution density of the cluster.

\begin{figure}
  \begin{tabular}{cc}
\includegraphics[width=0.5\textwidth]{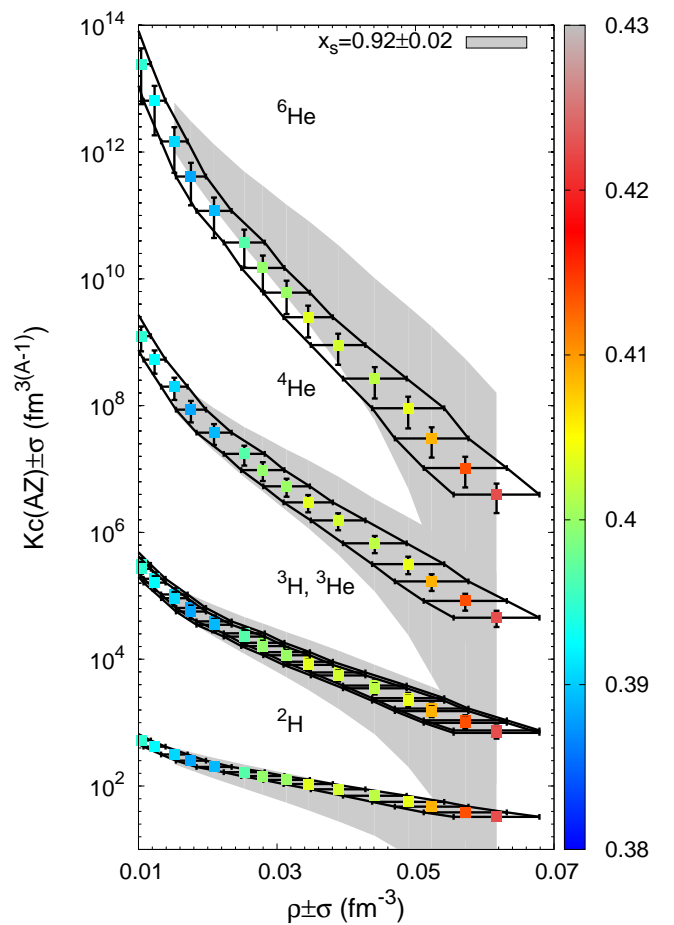} 
  \end{tabular}
\caption{(Color online) System $^{136}$Xe$+^{124}$Sn: The equilibrium constants as a function of the density. The full lines join the $1-\sigma$ uncertainty intervals. The grey bands are the equilibrium constants from a calculation \cite{PaisPRC2019} where we consider homogeneous matter with five light clusters, calculated at the average value of ($T$, $\rho_{\rm exp}$, $y_{pg_{\rm exp}}$), and considering cluster couplings in the range of  $ x_{s_i}=0.92\pm 0.02$. The color code represents the global proton fraction. } 
\label{fig3}
\end{figure}

The chemical equilibrium constants obtained with this model were compared with  the NIMROD results \cite{QinPRL108} obtained assuming an ideal gas expression for the determination of the nuclear density \cite{PaisPRC97,PaisPRC2019}, and a satisfactory agreement was  obtained for all clusters  but the deuteron using $x_s = 0.85 \pm 0.05$.

The prediction of this model is shown, for the thermodynamic conditions explored by the Xe$+$Sn systems, in the bottom panel of Fig.~\ref{fig1}. We can see that the calculation can reproduce the INDRA data, only if these latter are analyzed using the same hypotheses as in \cite{QinPRL108} (lower set of points). 
This suggests that the two sets are compatible, which points towards the validity of the statistical equilibrium hypothesis for both of them. However, it is also clear that the estimation  $x_s = 0.85 \pm 0.05$ overestimates the in-medium effects, once the consistent inclusion of the $C_{AZ}$ is accounted for.

To  estimate the effect of the correction, and, at the same time, determine the value of the in-medium parameter $x_s$ in a consistent way, we have compared the model of Refs.~\cite{PaisPRC97,PaisPRC2019} with this new analysis. 

In order to make this comparison,  we fix the temperature in each $(\rho_B,\ y_p)$ point by imposing that the isotopic thermometer $T_{HHe}$  evaluated in the  theoretical model, correctly reproduces the  measured $T_{HHe}$ value. A small difference between the input temperature of the theory, and the one estimated in the same calculation via the double ratios, is obtained, which does not exceed 10\% at the highest temperature. 
Indeed, the Albergo thermometer  \cite{AlbergoNCA89} used to estimate the temperature  is only valid under the assumption that the in-medium corrections to Eq.~(\ref{eq:spectra}), cancel in double isobar ratios, which is, in principle, not the case, if the correction does not scale linearly with the particle numbers. 
The resulting chemical constants are compared to the experimental ones in  Fig.~\ref{fig3}.  As we can observe,  the deuteron chemical constant behavior is now reproduced, and the chemical constants of $^3$He and $^3$H are almost superposed.  Very similar results are obtained for the other two experimental entrance channels (not shown).

In Refs.~\cite{PaisPRC97,PaisPRC2019},  we used $x_s=0.85$ in order to reproduce the results of  Qin \textit{et al.} \cite{QinPRL108}. With this improved analysis, a higher value $x_s>0.85$ is needed,
corresponding to smaller corrections and a larger dissolution density.
An optimal value can be extracted as $x_s=0.92\pm 0.02$. This value seems to reproduce reasonably well the whole set of experimental constants, and we have checked that it is still within the Virial EoS limits. This can be understood from the fact that the virial limit only concerns very low densities, where the predictions with the two different values of $x_s$ are very close (see Fig.~\ref{fig1}).
 
\begin{table}[htb]
\caption{\label{tab1} The experimental density and temperature from Ref.~\cite{QinPRL108}, and the theoretical calculation of the chemical equilibrium constant for the $\alpha-$particle, with two different scalar cluster-meson couplings.
}
\begin{ruledtabular}
\begin{tabular}{cccc}
 $\rho$ (fm$^{-3}$) & $T$ (MeV) &  $K_c$($^4$He) (fm$^9$) &  $K_c$($^4$He) (fm$^9$) \\
                               &                   &    $x_s=0.85$    &   $x_s=0.92$   \\
\hline
0.003	&	5.1	&	  0.22E+10 	&	 0.36E+10 	\\
0.005	&	5.6	&	  0.58E+09 	&	 0.12E+10 	\\
0.007	&	6.1	&	  0.20E+09 	&	 0.49E+09 	\\
0.009	&	6.5	&	  0.59E+08 	&	 0.19E+09 	\\
0.013	&	7.3	&	  0.14E+08 	&	 0.58E+08 	\\
0.015	&	7.8	&	  0.59E+07 	&	 0.29E+08 	\\
0.018	&	8.3	&	  0.24E+07 	&	 0.14E+08 	\\
0.021	&	9.0	&	  0.10E+07 	&	 0.69E+07 	\\
0.022	&	9.5	&	  0.62E+06 	&	 0.45E+07 	\\
0.025	&	10.0	&	  0.34E+06 	&	 0.27E+07 	\\
0.026	&	10.4	&	  0.23E+06 	&	 0.20E+07 
\end{tabular}
\end{ruledtabular}
\end{table}

The effect of the different estimation for the scalar coupling can be better appreciated from Table \ref{tab1}, which reports the predictions of the model for the $(\rho_B,T,y_p)$ trajectory estimated in Ref.~\cite{QinPRL108}, for which a large set of models and model assumptions was tested in Ref.~\cite{HempelPRC91}. We can see that, if we impose the consistent analysis of the INDRA data set as a new constraint, the theoretical model predictions for the chemical constants (last column) increase of a factor $\approx 1.5-10$, increasing with the density. This points towards smaller in-medium modifications than the ones extracted from the previous results in Ref.~\cite{HempelPRC91}.
 
In conclusion, a new analysis was performed based on INDRA data presented in Ref.~\cite{BougaultJPG19}.
We have shown that the presence of in-medium effects suppressing the cluster yields is necessary to explain the experimental data, giving rise to larger baryonic densities compared to the ideal gas limit.
The reduction factors were directly extracted from the data, under the unique condition that the different nuclear species in a given sample must  correspond to a unique common value for the density of the expanding source. 
We have verified that the three different data sets lead to fully compatible results for the corrections.
In the framework of a relativistic mean-field theoretical model \cite{PaisPRC97,PaisPRC2019}, these corrections can be interpreted as a stronger scalar meson coupling of the nucleons bound in clusters, which shifts the dissolution to higher densities.
 
In a future work, it would be extremely interesting to perform a new analysis of the experimental data of Ref.~\cite{QinPRL108}, with the same method as the one presented in this Letter, in order to check the consistency of the different data sets, and to settle the model dependence of the results.

\section*{ACKNOWLEDGMENTS}
This work was partly supported by the FCT (Portugal) Projects No. UID/FIS/04564/2019 and UID/FIS/04564/2020, and POCI-01-0145-FEDER-029912, and by PHAROS COST Action CA16214. We acknowledge support from R\'egion Normandie under RIN/FIDNEOS. H.P. acknowledges the grant CEECIND/03092/2017 (FCT, Portugal). For this work, she also acknowledges a PHAROS STSM grant and support from LPC (Caen). She is very thankful to F.G. and R.B. and the group at LPC (Caen) for the kind hospitality during her stay there.

\end{document}